\documentclass[prb,reprint, superscriptaddress, nofootinbib, amsmath,amssymb, aps,longbibliography]{revtex4-2}

\usepackage{amsfonts, amssymb, amsmath}
\usepackage{color}
\usepackage{comment}
\usepackage[normalem]{ulem}
\usepackage[english]{babel}
\usepackage[utf8]{inputenc}
\usepackage[colorinlistoftodos, color=green!40, prependcaption]{todonotes}
\usepackage{amsthm}
\usepackage{mathtools}
\usepackage{physics}
\usepackage{xcolor}
\usepackage{graphicx}
\usepackage[left=23mm,right=13mm,top=35mm,columnsep=15pt]{geometry} 
\usepackage{adjustbox}
\usepackage{placeins}
\usepackage[T1]{fontenc}
\usepackage{lipsum}
\usepackage{csquotes}
\usepackage{float}
\usepackage[pdftex, pdftitle={Article}, pdfauthor={Author}]{hyperref} 

\usepackage{amsfonts}           
\usepackage{dsfont}             %
\usepackage{braket} 
\usepackage{mathrsfs}           
\usepackage{bm}                 
\usepackage{nccmath}
\usepackage{mathtools}
\usepackage{lipsum}
\usepackage{svg} 
\usepackage{multirow}

\begin{document}
\title{Time-dependent fluctuating local field approach for description of the correlated fermions dynamics}

\author{L.~D. Silakov}
\affiliation{Russian Quantum Center, Moscow 121205, Russia}
\affiliation{Moscow Institute of Physics and Technology, Dolgoprudny, 141701, Russia}
\author{Ya.~S. Lyakhova}
\affiliation{Russian Quantum Center, Moscow 121205, Russia}
\author{A.~N. Rubtsov}\email{ar@rqc.ru}
\affiliation{Russian Quantum Center, Moscow 121205, Russia}
\affiliation{Lomonosov Moscow State University, Moscow 119991, Russia}

\date{\today} 

\begin{abstract}
We formulate a time-dependent Fluctuating Local Field (TD-FLF) method for correlated fermion dynamics, extending the stationary FLF approach. The wavefunction is approximated as an ensemble of non-interacting states subject to a classical fluctuating field, with dynamics encoded in the field’s time-dependent distribution. This reduces the time-dependent Schr\"odinger equation to a generalized eigenvalue problem in a significantly reduced basis. Applied to half-filled 2D Hubbard lattices, TD-FLF yields highly accurate results, outperforming mean-field theory and capturing oscillation frequencies and amplitudes in good agreement with exact diagonalization. Its low computational cost and flexibility make TD-FLF a promising tool for simulating driven correlated systems.
\end{abstract}


\maketitle

\section{Introduction} \label{sec:introduction}

It is hard to overestimate the importance of problems related to the description of correlated electron dynamics. This topic plays a crucial role in multiple fields, including electronic phase transitions \cite{NatCommPhaseTrans}, quantum chemistry \cite{SciRep2020}, and the behavior of fermions in optical lattices \cite{PhysRevLett.94.080403}. Recent advances in time-resolved spectroscopy and cold-atom technologies have paved the way for experimental investigations of real-time quantum dynamics \cite{Sponselee_2019}.

All of the above has spurred significant interest in the theoretical description of such processes. Moreover, recent advances in simulating quantum dynamics on quantum processors have further intensified this interest. The community regards real-time simulations of correlated electrons as one of the most significant—and practically relevant—tasks for quantum computers in the coming years  \cite{doi:10.1021/acs.accounts.1c00514}. Various approaches have been proposed to derive quantum algorithms capable of capturing the dynamics of correlated fermions while remaining tractable on quantum simulators \cite{PhysRevResearch.7.043018,PhysRevB.111.144306,zmwm-gdmw}. And there are already promising experimental results for some chemical dynamics \cite{jacs.5c03336}. Recently, scientists from Phasecraft managed to exactly simulate a 28-sites Hubbard system on Quantinuum’s System Model H2-2 trapped-ion quantum computer \cite{alam2025fermionicdynamicstrappedionquantum}. The authors obtained the time evolution of physical observables, including the local charge density, spin correlation functions, and doublon number. We emphasize that exact classical simulation of spinful, correlated systems of this size lies beyond current computational capabilities. The Phasecraft team notes that their results differ from those obtained via classical methods, such as the time-dependent variational principle and Majorana propagation, making the findings particularly intriguing. This discrepancy both motivates further development of quantum hardware and underscores the need to improve classical computational methods to challenge quantum computers. Given that the computational cost of exact real-time simulation grows exponentially with the number of degrees of freedom, approximate methods inevitably play a central role here.

A number of computational methods have been developed to approximately describe the dynamics of correlated quantum systems. Most of these constitute time-dependent (TD) extensions of established stationary approaches. For instance, the mentioned time-dependent variational principle (TDVP) employs parametrized ansatz wave functions to evolve quantum states in time \cite{PhysRevB.55.8226,PhysRevB.83.165105}. Variants of this framework differ primarily in the choice of ansatz. In Ref. \cite{PhysRevB.55.8226}, the authors use generalized coherent states to formulate semi-classical equations of motion for the Hubbard model, yielding rich physical insights, albeit at the cost of requiring detailed knowledge of the Hamiltonian’s dynamical algebra. In contrast, Ref. \cite{PhysRevB.83.165105} applies the time-dependent Gutzwiller approximation to study quench dynamics in the Hubbard model. Beyond TDVP-based schemes, more sophisticated methods have been proposed, including time-dependent density functional theory (TDDFT) \cite{annurev-physchem-082720-124933}, hybrid TDDFT + dynamical mean-field theory (TDDFT+DMFT) \cite{PhysRevLett.106.116401}, time-dependent numerical renormalization group (TD-NRG) \cite{PhysRevB.90.035129}, and time-dependent Monte Carlo (TD-MC) \cite{TDMC2021}. Among all of these, the multiconfigurational time-dependent Hartree–Fock (MCTDHF) method stands out as one of the most powerful tools for real-time simulations in quantum chemistry \cite{PhysRevA.71.012712}. Its central idea is to express the wave function as a time-dependent superposition of Slater determinants built from time-dependent orbitals. The Dirac–Frenkel time-dependent variational principle is then employed to derive coupled equations of motion for both the superposition coefficients and the orbital wave functions. The MCTDHF framework and its extensions have been successfully applied to simulate coupled electronic and nuclear dynamics in molecules \cite{PhysRevA.83.063416}, and can be generalized to bosonic systems as well \cite{1.2771159}.

Further development of this broad family of TD methods requires addressing their limitations. Arguably, the most universal constraint is the limited time span accessible in simulations. In addition, one must contend with difficulties inherited from the underlying stationary schemes, most notably, the rapid (often exponential) growth of computational cost with system size and complexity.

In this work, we present a TD extension of the recently introduced stationary fluctuating local field (FLF) approach \cite{Rubtsov2018FLF}. At the heart of this method lies the assumption that correlation effects can be captured by a small number of collective fluctuation modes. Practically, the FLF approach builds upon a simple parental approximation, in which an order parameter is held fixed, and restores its fluctuations by coupling it to an auxiliary, artificially introduced fluctuating field. In this way, the FLF scheme retains the low computational cost of the parental approximation while avoiding its unphysical consequences \cite{rubtsov2020collective,lyakhova2022fluctuatingMolecules,PhysRevB.110.245134}.


To formulate the TD fluctuating local field (TD-FLF) approach, we begin by representing the many-body state as an ensemble of single-particle states, each corresponding to a distinct value of an external field. To capture the system’s dynamics, we postulate that all time dependence resides in the real-valued distribution function over this ensemble. We demonstrate that, under this ansatz, the time-dependent Schr\"odinger equation reduces to a generalized eigenvalue problem for the distribution function. Notably, the dimensionality of this problem is substantially smaller than that of the exact Fock-space representation. Moreover, the computational cost of the TD-FLF scheme increases only modestly with growing system complexity. To illustrate this, consider again MCTDHF method: it requires evaluating a moderate number of Slater determinants, each built from different time-dependent orbitals --- an aspect that drives up computational effort. In contrast, TD-FLF also employs Slater determinants constructed from single-particle states, and while their number may be large (yet moderate, thanks to the reduced basis), all share the same set of orbitals. This key distinction suppresses the rapid growth of numerical cost typically encountered in other correlated methods.


We test our TD-FLF approach on the description of magnetic properties of the two-dimensional Hubbard lattice at half filling. This system is known to exhibit strong antiferromagnetic (AFM) fluctuations, even in the weak- to moderate-coupling regimes \cite{PhysRevB.65.081105}, making its theoretical description nontrivial. To enable rigorous benchmarking, we consider small plaquettes, for which exact diagonalization (ED) results are available \cite{PhysRevB.46.11779,doi:10.1142/S0217979289001202}. Numerical simulations of the magnetization dynamics show that the TD-FLF method reproduces the ED results exactly for the $2\times 2$ lattice and provides a highly accurate approximation for the $2\times 4$ lattice, substantially outperforming the parental mean-field approximation in both cases. Notably, our calculations show that TD-FLF has an advantage over more sophisticated methods like TDDFT-DMFT \cite{PhysRevLett.106.116401} by successfully capturing damping of magnetization oscillations, which manifests influence of interaction on the correlated fermions dynamics. A subsequent Fourier analysis of the time-dependent magnetization reveals a characteristic oscillatory pattern; remarkably, the TD-FLF scheme captures both the frequencies and relative amplitudes of these oscillations quite accurately.

\section{Stationary FLF approach} \label{sec:MF}
The aim of this work is to construct time-dependent version of the FLF approach. Thus we start with the general remind of the stationary FLF scheme.

The FLF approach is based on the assumption, that correlation effects are largely associated with fluctuations of some order parameter. This assumption dictates three main steps of the FLF approach, they are: (i) definition of the leading fluctuation channel; (ii) introduction of the auxiliary fluctuating classical field in this channel; (iii) integrating over all values of this field. These points can be expressed via the exact transformation of the partition function
\begin{align}
\label{eq:FLF_transform}
\nonumber
    \mathcal{Z} &= \int e^{-\mathcal{S}_0[c^*,c] - \mathcal{S}_U[c^*,c]} \mathcal{D} \left[c^*, c \right]\\
    \nonumber
    &=\left(\frac{\beta N}{2\pi \lambda}\right)^{3/2} \iint  e^{-\mathcal{S}_0[c^*,c] - \mathcal{S}_U[c^*,c]-\frac{\beta N}{2 \lambda} \left(\vec{\nu} - \vec{\mathcal{O}} \right)^2} \mathcal{D}[c^*,c] \, d \vec{\nu}\\
    &=\left(\frac{\beta N}{2\pi \lambda}\right)^{3/2} \int \mathcal{Z}_\nu e^{-\frac{\beta N}{2 \lambda} \nu^2}d\vec{\nu}.
\end{align}
Here $-\mathcal{S}_0[c^*,c] - \mathcal{S}_U[c^*,c]$ is the exact action including free part $\mathcal{S}_0$ and interaction $\mathcal{S}_U$ and expressed in terms of Grassmann variables $c^*,\,c$, $\vec{\nu}$ is the fluctuating field coupled to the order parameter $\vec{\mathcal{O}}$, and
\begin{equation}
    \mathcal{Z}_\nu = \int e^{-\mathcal{S}_0[c^*,c] - \mathcal{S}_U[c^*,c] + \beta N \left(\vec{\nu}\vec{\mathcal{O}}\right) - \frac{\beta N \left(\vec{\mathcal{O}} \vec{\mathcal{O}}\right)}{2\lambda}} \mathcal{D}[c^*,c]
\end{equation}
is the exact partition function of the initial system exposed to external field $\vec{\nu}$ and with effective interaction term $\frac{\beta N \left(\vec{\mathcal{O}} \vec{\mathcal{O}}\right)}{2\lambda}$. Free parameter of the stationary FLF scheme $\lambda$, and the pre-integral multiplier that makes this transformation exact are not relevant for the current work.

In practice, exact partition function $\mathcal{Z}_\nu$ in \eqref{eq:FLF_transform} is replaced with an approximate one. The simplest choice here is the mean-field (MF) system. It means that we approximate single interaction-full system with an ensemble of free systems embedded in external fluctuating field.

For concreteness, we consider the example of the two-dimensional Hubbard model
\begin{equation}\label{eq:Exact_Hamiltonian}
    \begin{split}
    \hat{H} =& -V\sum_{\langle j,j' \rangle, \sigma} \left( \hat{c}_{j, \sigma}^{\dag}\hat{c}_{j', \sigma}^{} + \hat{c}_{j', \sigma}^{\dag}\hat{c}_{j, \sigma}^{} \right)\\
    &+ U\sum_{j}\left(\hat{n}_{j\uparrow}-\frac{1}{2}\right) \left(\hat{n}_{j\downarrow}-\frac{1}{2}\right) - \sum_j\vec{h}_j\hat{\vec{S}}_j.
    \end{split}
\end{equation}
Here, the $\hat{c}_{j \sigma}^{(\dag)}$ operators correspond to annihilation (creation) of fermions, where the subscripts denote the position $j$ and spin projection $\sigma \in \{\uparrow, \downarrow\}$. $V$ is the hopping amplitude between two nearest-neighbor sites $\langle j,j' \rangle$. $U$ is the on-site Coulomb repulsion interaction between fermionic densities $\hat{n}_{j \sigma} = \hat{c}^{\dag}_{j \sigma} \hat{c}_{j\sigma}$ with opposite spin projections. $\vec{h}_j$ is the external magnetic field on the $j$-site coupled to the local spin $\hat{\vec{S}}_j=\hat{c}^\dag_{j\sigma}\vec{\sigma}_{\sigma\sigma'}\hat{c}_{j\sigma'}e^{i\bm{Kr}_j}$ with $\bm{K}=(\pi,\pi)$, and $\bm{r}_j$ points the position of the $j$-site.

This system, being half-filled, is known to possess strong fluctuations in the anti-ferromagnetic (AFM) spin-channel, which is characterized by the collective order parameter $\hat{\vec{S}} = \frac{1}{N}\sum_j\hat{\vec{S}}_j$. This property manifests in fiction Neel point predicted by the mean-field (MF) approximation. In this case, we can take $\vec{\mathcal{O}}\equiv \hat{\vec{S}}$ in \eqref{eq:FLF_transform}, and approximate $\mathcal{Z}_\nu$ with the MF partition function. It was shown, that such an approach allows to cure fiction N\'eel transition, and obtain qualitatively and quantitatively good results in comparison with much more sophisticated and numerically consuming methods \cite{rubtsov2020collective,PhysRevB.110.245134}. Here we are to exploit the main idea of FLF to describe time dynamics of correlated fermions.

\section{Time-dependent FLF approach}

\subsection{Time-dependent extension of the FLF scheme}
The time-dependent FLF scheme, which we aim to develop, should allow to compute the dynamics of correlated system. Originally it can be described by the TD Schrödinger equation (TDSE) in terms of the quantum state of the system $\ket{\psi(t)}$:
\begin{equation}\label{eq:Schordinger_equation}
    i\partial_t \ket{\psi(t)} = \hat{H}\ket{\psi(t)},
\end{equation}
where $\hat{H}$ is the Hamiltonian. Here and in the following we use the natural units system: $\hbar = 1$. Acting in the spirit of stationary FLF approach, we approximate this state as an ensemble of non-interacting MF systems' eigenstates $\ket{\varphi^{m}_{\vec{\nu}}}$ corresponding to different values of external field $\vec{\nu}$:
\begin{equation}\label{eq:FLF_definition}
    \ket{\psi(t)} \approx \ket{\psi^{FLF}(t)} = \sum_{m} \int  f^{m}(\vec{\nu};t)\ket{\varphi^{m}_{\vec{\nu}}} \ d\vec{\nu}.
\end{equation}
Here $m$ is the number of eigenstate, and $f^m(\vec{\nu};t)$ is a distribution function over these states. We will omit the index $m$ and the corresponding summation in the further analytical derivation for brevity. Note, that introduction of function $f(\vec{\nu};t)$ is the generalization of \eqref{eq:FLF_transform}, where we assumed it to be Gaussian. In the present work we show, that this distribution can be explicitly found.

It is clear from \eqref{eq:FLF_definition}, that in our scheme we ascribe time-dependence of the quantum state to the distribution function. It allows to reduce the TDSE to the equation on $f(\vec{\nu};t)$:
\begin{equation}
\label{eq:TDSE_f}
    i \mathds{1}_{\nu'\nu}\partial_t f(\vec{\nu};t) = H_{\nu'\nu}f(\vec{\nu};t),
\end{equation}
where $\mathds{1}_{\nu'\nu} \equiv \braket{\varphi_{\vec{\nu}'}|\varphi_{\vec{\nu}}}$, and $H_{\nu'\nu}\equiv\braket{\varphi_{\vec{\nu}'}|\hat{H}|\varphi_{\vec{\nu}}}$.

\subsection{Basis reduction}
\begin{figure}[]
    \centering 
    \includegraphics[width=0.5\textwidth]{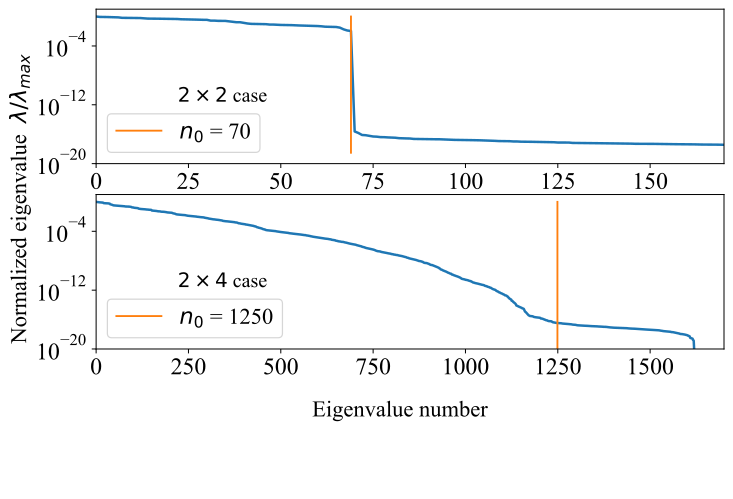}
    \vspace{-16mm}
    \caption{Spectrum of matrix $\mathds{1}$ eigenvalues $\lambda$ divided by the maximum eigenvalue $\lambda_{max}$ and the size $n_0$ of effective basis for $2\times 2$ and $2\times 4$ half-filled Hubbard lattices.}
    \label{fig:Spectrum}
\end{figure}

On practice, FLF expansion \eqref{eq:FLF_definition} is over discrete values of $\vec{\nu}$, and thus $\mathds{1}_{\nu'\nu}$ and $H_{\nu'\nu}$ represent matrices. The matrix $\mathds{1}_{\nu'\nu}$ though is not unity in general, as $\{\ket{\varphi_{\vec{\nu}}}\}$ are not necessarily orthogonal. This peculiarity can be cured by diagonalizing $\mathds{1}$ and keeping only those eigenstates $\{\vec{\eta}_j\}$, which correspond to significantly non-zero eigenvalues. Fig. \ref{fig:Spectrum} represents the results of such calculation on the example of the spectrum of $\mathds{1}$ for $2\times 2$ and $2\times 4$ size Hubbard lattices. One can see that the eigenvalues $\lambda/\lambda_{max}$ decrease rapidly after some number $n_0$, which we fix as the reduced basis size. In what follows we will call this basis \textit{effective}.

\begin{figure}[]
    \centering 
    \includegraphics[width=0.5\textwidth]{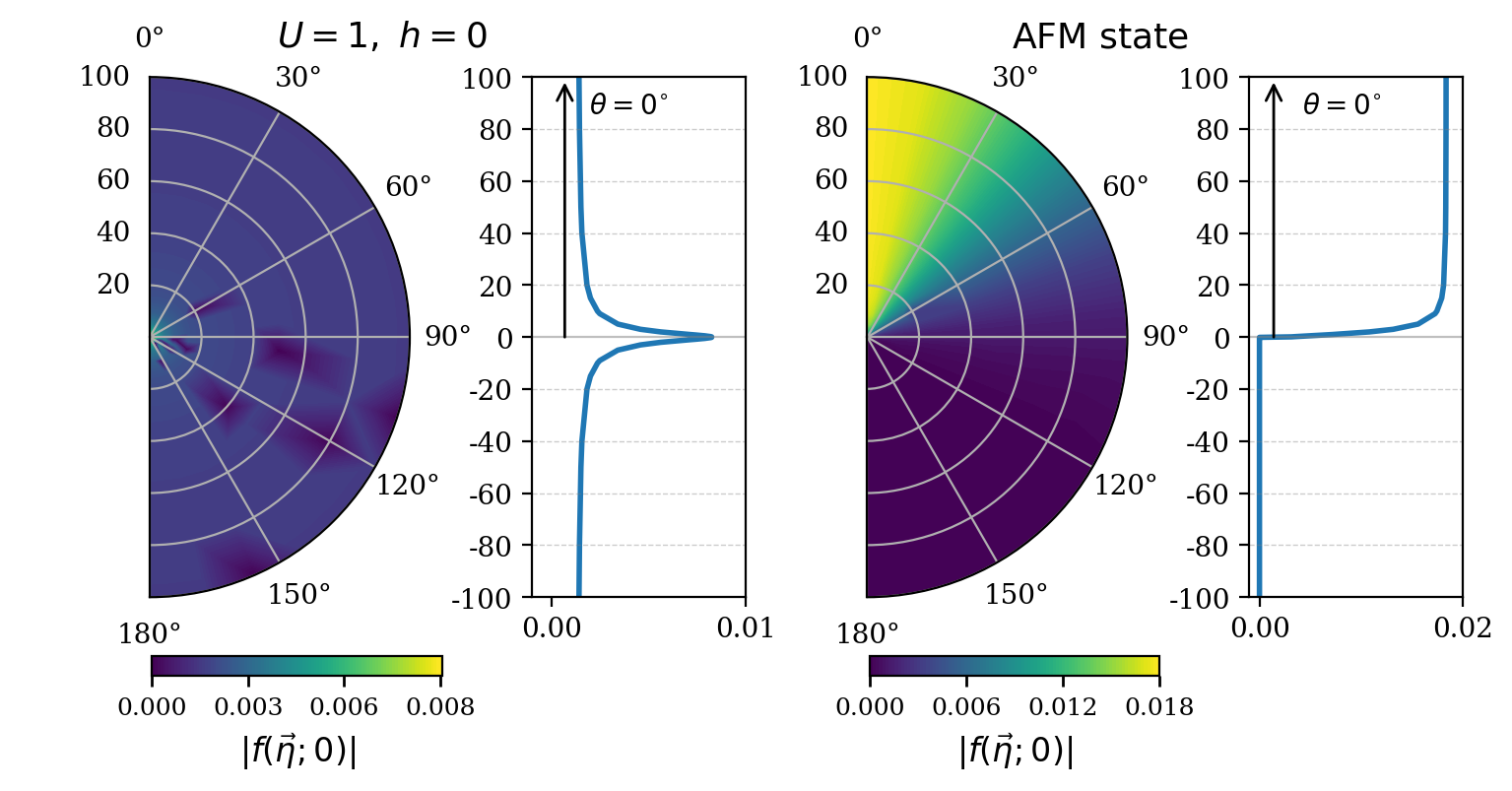}
    \vspace{-8mm}
    \caption{Distribution function $|f(\vec{\eta};0)|$ for the ground state at $U=1$, $h=0$ (left panel), and for the AFM state (right panel) obtained as expansion over the mean-field basis in $2\times4$ case. The polar plots show $|f(\vec{\eta};0)|$ as a function of the field magnitude $|\vec{\eta}|$ and direction-defining angle $\theta$; the side panels display the profile at $\theta = 0$.}
    \label{fig:Distribution}
\end{figure}
It worth noting that the use of the reduced effective basis is one of the key points making the TD-FLF scheme highly numerically effective. This is because the effective basis dimension is usually significantly lower than in the original system. In the case of $2\times 2$ system we obtain 70 effective basis states against 256 states in the full Fock basis. For the $2\times 4$ this difference is even more. Here we have 1250 effective basis states against 65536 states in the full Fock basis. In order to evaluate the quality of such approximation we calculate the ground state energy of two-dimensional Hubbard lattice of different sizes and at different values of $U$. The results, obtained within FLF and its parental MF approximation, were compared to ED data, and are listed in Table \ref{tab:E_0_values}. We see that in all cases FLF significantly overcome the MF and are in good accordance with the exact results. It allows us to consider the chosen reduced bases as satisfactory.

\begin{table*}[t]
\centering
\setlength{\tabcolsep}{8pt} 
\begin{tabular}{ |c|c|c|c|c|c|c| } 
\hline
\multirow{2}{*}{$U/V$} 
    & \multicolumn{3}{c|}{$2\times2$ lattice} 
    & \multicolumn{3}{c|}{$2\times4$ lattice} \\
\cline{2-7}
    & Exact & FLF & MF & Exact & FLF & MF \\
\hline
0   & -9.062  & -9.062  & -9.062  & -17.308 & -17.308 & -17.308 \\
0.5 & -9.229  & -9.229  & -9.225  & -17.496 & -17.488 & -17.483 \\
1.0 & -9.411  & -9.411  & -9.398  & -17.734 & -17.704 & -17.685 \\
2.0 & -9.824  & -9.824  & -9.779  & -18.359 & -18.259 & -18.197 \\
4.0 & -10.828 & -10.828 & -10.698 & -20.168 & -19.913 & -19.771 \\
8.0 & -13.429 & -13.429 & -13.193 & -25.435 & -25.096 & -24.952 \\
16.0& -20.019 & -20.019 & -19.828 & -39.048 & -38.840 & -38.764 \\
\hline
\end{tabular}
\caption{Ground state energy of $2\times2$ and $2\times4$ half-filled Hubbard lattices embedded in $h=0.5$ magnetic field.}
\label{tab:E_0_values}
\end{table*}

Rewriting \eqref{eq:TDSE_f} in the effective basis, we obtain the final equation on the distribution function $f(\vec{\eta};t)$
\begin{equation}
    i\partial_tf(\vec{\eta};t) = H_{\eta'\eta}f(\vec{\eta};t).
\end{equation}
Solving this equation with some given initial state $f(\vec{\eta};0)$ allows to construct the full FLF state \eqref{eq:FLF_definition}, and thus calculate physical observables. Figure \ref{fig:Distribution} represents plots for two different distribution functions $f(\vec{\eta};t=0) = \langle \varphi_{\vec{\eta}} | \psi(t=0) \rangle$. As the FLF scheme is based on the assumption that collective fluctuations dominate the physics of correlated fermions, the initial state should be chosen consistently with the leading fluctuation channel. Thus, for half-filled Hubbard lattice, which is known to possess strong AFM fluctuations, it is reasonable to chose fully AFM-ordered initial state. For describing, say, hole-doped systems in this approach one would need more complex initial state being the superposition of different states with one hole; and/or superposition of differently ordered states chosen in accordance with partial nesting observed for such systems.

\section{Numerical results and discussion} \label{sec:Num_res}
As a concrete example, we consider how the staggered magnetization of the half-filled Hubbard lattice evolves over time: 
\begin{equation}\label{eq:observable_value}
    \braket{S}(t) = \frac{1}{N}\sum_j \braket{\psi(t)|\hat{n}_{j\uparrow} - \hat{n}_{j\downarrow}|\psi(t)} e^{i\bm{Kr}_j}.
\end{equation}
We start with the initial fully AFM-ordered state, which means the initial magnetization is $\braket{S}(t=0) = 1$. Again, we consider small lattices of size $2\times 2$ and $2\times 4$. It simultaneously solves two problems. At first, it allows us to directly access exact reference results at low computational cost. But what is more important, it eliminates the need to take into account spatial inhomogeneities of collective fluctuations, which are proven to evolve the physics of system under consideration \cite{lyakhova2022fluctuatingMolecules}. It is reasonable to expect, that temporal evolution of collective fluctuations involves small lattices as a whole, thus revealing its complicated pattern rapidly.

\begin{figure}[H]
\centering
\includegraphics[width=0.48\textwidth]{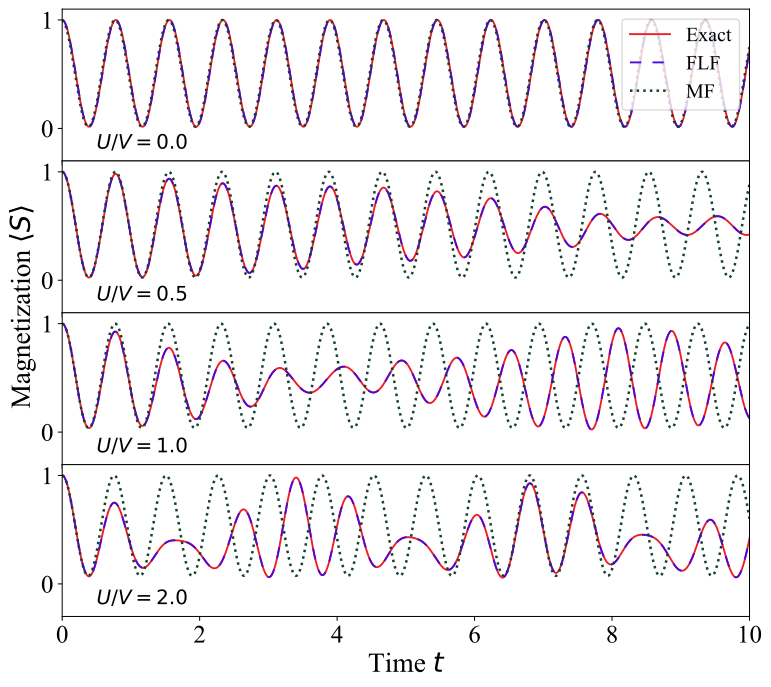}
\vspace{-3mm}
\caption{Evolution of magnetization of $2\times 2$ half-filled Hubbard lattice embedded in $h=0.5$ magnetic field. MF approximation and TD-FLF results are compared with the numerically exact (Exact) reference data.}
\label{fig:Exact_MF_FLF_2x2}
\end{figure}

\begin{figure}[H]
\includegraphics[width=0.48\textwidth]{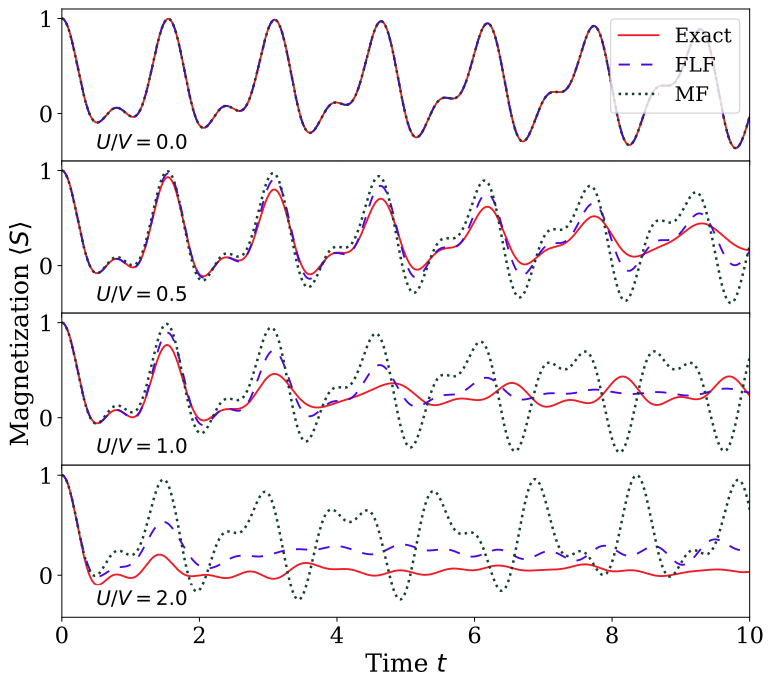}
\vspace{-3mm}
\caption{Evolution of magnetization of $2\times 4$ half-filled Hubbard lattice embedded in $h=0.5$ magnetic field. MF approximation and TD-FLF results are compared with the numerically exact (Exact) reference data.}
\label{fig:Exact_MF_FLF_2x4}
\end{figure}

The numerical results, obtained within TD-FLF, MF and ED approaches for half-filled Hubbard lattices, are presented on the Figures \ref{fig:Exact_MF_FLF_2x2} and \ref{fig:Exact_MF_FLF_2x4}. First, we see that the larger $U$ is, the worse results gives MF approximation. TD-FLF in contrast catches changes is the temporal fluctuation pattern of the magnetization. For $2\times 2$ system TD-FLF totally coincides with ED, which is reasonable, as the size $70$ of the effective basis in this case is exactly the same as the full basis for $2\times 2$ system at half-filling with equal number of spin-up and spin-down fermions. For $4\times 2$ TD-FLF allows to capture diminishing of the fluctuations amplitude and their general pattern. We also note that at short time span TD-FLF reproduces exact results very closely.

A significant feature of plots on Figure \ref{fig:Exact_MF_FLF_2x4} concerns the decrease of oscillations amplitude. The stronger the interaction $U$, the more rapid is damping. It thus reveals the Mott-type correlation of fermions. For a quantitative characterization of this behavior we fit the TD-FLF and exact data with an exponential decay $\propto e^{-\xi t}$ for $2\times 4$ lattice at $U/V = 1$. It supplied us with the damping rate estimation, which appears to be $\xi_{ED} \approx 0.24$ and $\xi_{TD-FLF} \approx 0.25$ for exact and TD-FLF data respectively.

\begin{figure}
\includegraphics[width=0.48\textwidth]{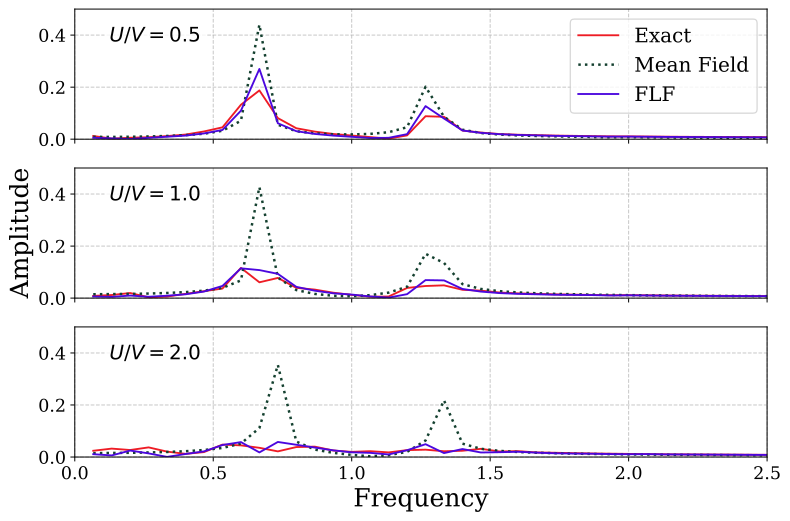}
\vspace{-3mm}
\caption{Fourier spectrum of temporal magnetization oscillations of 2 $\times$ 4 Hubbard lattice. Mean-field  approximation (MF) and fluctuating local field method (FLF) results are compared with the numerically exact (Exact) reference data.}
\label{fig:Fourier_MF_FLF_2x4}
\end{figure}

Finally, we considered  the pattern of magnetic temporal oscillations we observe. With this aim we numerically apply the Fourier transform to \eqref{eq:observable_value}. The results are presented on the Figure \ref{fig:Fourier_MF_FLF_2x4}. One can see that TD-FLF scheme effectively captures the frequencies and amplitudes of magnetic oscillations, and thus can be used as a good instrument for studying the dynamical ordering patterns of the system.

\section{Conclusion} \label{sec:Conclusion}


In this work, we have derived a time-dependent version of the fluctuating local field (FLF) approach. The original method was recently introduced for the description of correlated fermionic systems. The stationary FLF scheme has proven to be a cost-effective and flexible numerical method for systems in which fluctuations of collective order parameters play a dominant role in the underlying physics.


Here, we exploit the flexibility of the original FLF approach to develop its time-dependent (TD) version. To this end, we approximate the full time-dependent wave function by an ensemble of non-interacting ones, each corresponding to a distinct value of an external fluctuating field. This reduces the time-dependent Schr\"odinger equation to an evolution equation for the field’s distribution function. We further propose an efficient scheme to reduce the number of non-interacting states required in the calculation. Altogether, these features make the TD-FLF approach exceptionally lightweight in terms of numerical resources.


To verify the applicability of our method, we examined the temporal dynamics of magnetization in the Hubbard model, a system known to exhibit strong antiferromagnetic fluctuations at half-filling. In our calculations, we focused on small lattices. It allowed us to use exact diagonalization technique to obtain reference data, and eliminated possible inhomogeneity effects. We demonstrated that TD-FLF enables simulations extending significantly further in time than the parental mean-field approximation, across weakly ($U/V = 0.5$) and moderately ($U/V = 1,\,2$) correlated regimes. TD-FLF scheme has also proven to be sensitive enough to correctly capture damping of magnetization oscillations, which is caused by electron-electron interaction. In this regard TD-FLF scheme exceeds mode sophisticated and computationally consuming methods such as TDDFT-DMFT \cite{PhysRevLett.106.116401}. Fourier analysis of these results further highlights the advantage of TD-FLF, revealing that our approach accurately describes both the frequencies and amplitudes of elementary oscillations in close agreement with exact numerical data.


Although in this work we tested TD-FLF on one-orbital half-filled Hubbard lattices with moderate interactions, the approach has already proved to be a promising numerical scheme, which combines flexibility and computational efficiency. It is readily extensible to incorporate multiple fluctuation channels, spin-imbalanced systems, multi-orbital case, finite temperatures, and beyond. It is also important to note, that the TD-FLF approach is not limited to weakly and moderately interacting systems. The interaction strength should be taken into account by the parental approximation in \eqref{eq:FLF_definition}. For instance, the case $U/V \gg 1$ would demand one-site (atomic) approximation as the parental.

In an era of rapidly advancing experimental techniques and quantum processing hardware, the demand for theoretical tools that deliver not only quantitative predictions but also physical insight into correlated systems has become particularly acute. We hope that the TD-FLF framework presented here will serve as a foundation for such a powerful and versatile method.

\section*{Acknowledgments} \label{sec:acknowledgments}

We acknowledge the support from the Russian Quantum Technologies Roadmap.

\bibliographystyle{apsrev}
\bibliography{mybib}

\end{document}